\documentclass[twocolumn,amsmath,amssymb,prc,superscriptaddress,floatfix,
showpacs]{revtex4}
\usepackage{graphicx}
\usepackage{subfigure}
\usepackage{color}
\usepackage{epsfig}
\usepackage[T1]{fontenc} 

\graphicspath{{.}{./tovfigures/}}

\begin{document}
\title{On the appearance of hyperons in neutron stars}

\author{H.~\DJ apo}
\email[E-Mail:]{haris@crunch.ikp.physik.tu-darmstadt.de}
\affiliation{Institut
  f\"ur Kernphysik, TU Darmstadt, Schlo{\ss}gartenstr.~9, D-64289 Darmstadt,
Germany}
\author{B.-J. Schaefer}
\email[E-Mail:]{bernd-jochen.schaefer@uni-graz.at}
\affiliation{Institut f\"ur Physik, Karl-Franzens-Universit\"at Graz,
  Universit\"atsplatz 5, A-8010 Graz, Austria} 
\author{J. Wambach} 
\affiliation{Institut
  f\"ur Kernphysik, TU Darmstadt, Schlo{\ss}gartenstr.~9, D-64289 Darmstadt,
Germany}
\affiliation{Gesellschaft
  f\"ur Schwerionenforschung mbH, Planckstr.~1,  D-64291 Darmstadt, Germany}

\date{\today}

\begin{abstract}
  By employing a recently constructed hyperon-nucleon potential the
  equation of state of $\beta$-equilibrated and charge neutral
  nucleonic matter is calculated. The hyperon-nucleon potential is a
  low-momentum potential which is obtained within a renormalization group
  framework. Based on the Hartree-Fock approximation at zero temperature
  the densities at which hyperons appear in neutron stars are
  estimated. For several different bare hyperon-nucleon potentials and
  a wide range of nuclear matter parameters it is found that hyperons
  in neutron stars are always present. These findings have profound
  consequences for the mass and radius of neutron stars.
\pacs{13.75.Ev,
21.65.Mn, 
26.60.-c
}
\end{abstract}

\maketitle

\section{Introduction}
\label{sec:intro}

Neutron stars (NS) are compact objects with interior densities of
several times normal nuclear density. The precise and detailed
structure and composition of the inner core of a NS is not known 
at present. Several possibilities such as mixed phases of quark
and nuclear matter, kaon or pion condensates or  color superconducting 
quark phases are under debate.

In the present work the influence of hyperons with strangeness $S=-1$
($\Lambda$, $\Sigma^-$, $\Sigma^0$ and $\Sigma^+$) on the composition
and structure of a NS is investigated. In this context the central and
essential quantity to be analyzed is the equation of state (EoS). The
EoS determines various NS observables such as the mass range or the
mass-radius relation of the star. The composition and structure of
neutron and also of protoneutron stars have been investigated in
detail with a wide range of EoS for dense nuclear matter
\cite{Prakash:1996xs,Strobel:1999vn}. The emergence of hyperons for
increasing nucleon densities has been suggested in the pioneering work
of \cite{1960SvA.....4..187A}. Since then the impact of hyperons on
dense matter has been studied extensively with different approaches,
see e.g.~\cite{Pandharipande:1971up,Glendenning:1984jr,
  Keil:1995hw,Schaffner:1995th,Prakash:1996xs,Baldo:1999rq,Vidana:2000ew}
Unfortunately, the details of the hyperon-nucleon ($YN$) interaction
and even more of the hyperon-hyperon ($YY$) interaction are known only
poorly. The limited amount of available experimental data enables the
construction of many different potentials. For example, the Nijmegen
group has proposed six different potentials which all describe the
low-energy data such as phase shifts equally well, see e.g.~\cite{rijken}.

We wish to explore the differences between the available $YN$
interactions and their influences on the appearance of hyperons in
neutron stars. For this purpose the hyperon $YN$ threshold densities
are calculated. Since the $\Sigma^0$ and $\Sigma^+$ hyperons are
heavier than the $\Lambda$ and $\Sigma^-$ hyperons they typically
appear later. Hence, we will focus on the $\Lambda$ and $\Sigma^-$
hyperons in the following.

The paper is organized as follows: In Sec.~\ref{sec:EoS} the EoS of
dense matter including hyperons are calculated. For the pure nucleonic
part of the EoS a parameterization is used. A central quantity which
enters the EoS is the single-particle potential. The derivation of the
single-particle potential for hyperons is given in Sec.~\ref{sec:SPP}.
The requirements which are necessary for equilibrium are discussed in
Sec.~\ref{sec:beta} and the results are collected in
Sec.~\ref{sec:Threshold densities}. The next section
Sec.~\ref{sec:structure} shows the calculations of neutron stars with
and without hyperons. Finally, the work is summarized and conclusions
are drawn in Sec.~\ref{sec:summary}.

\section{Single-particle potentials}
\label{sec:SPP}

Single-particle potentials are a useful and important tool to
determine the density at which hyperons begin to appear in baryonic
matter. They can be obtained from an effective low-momentum $YN$
potential $\ensuremath{V_\text{low k}}$ \cite{schaefer, wagner} in the
Hartree-Fock approximation. Details concerning the derivation of the
$YN$ single-particle potentials in this approximation can be found in
\cite{Djapo:2008qv}. With these single-particle potentials the chemical
potentials and particle energies in $\beta$-equilibrated matter can be
calculated. This allows to compute the threshold densities for the
appearance of a given hyperon species and to establish the
concentrations of all particles in dense matter at a given density. In
the process we will also determine the EoS for the mixture of leptons
and baryons (electrostatic interactions are neglected because their
energies are orders of magnitude smaller than the other interaction
energies).

Our many-body scheme employs several ``bare'' $YN$ interactions as
input for the $\ensuremath{V_\text{low k}}$ calculation: the original
Nijmegen soft core model NSC89 \cite{nsc89}, a series of new soft core
Nijmegen models NSC97a-f \cite{rijken}, a recent model proposed by the
J\"ulich group J04 \cite{juelich} and chiral effective field theory
(\ensuremath{{\chi\text{EFT} }}) \cite{poli}. The first three models
are formulated in the conventional meson-exchange (OBE) framework,
while the \ensuremath{{\chi\text{EFT} }} is based on chiral
perturbation theory (for recent reviews see
e.g.~\cite{bedaque-2002-52, epelbaum-2006-57,Furnstahl:2008df}). In
the \ensuremath{{\chi\text{EFT} }} approach a cutoff
$\Lambda_\ensuremath{{\chi\text{EFT} }}$ enters which we fix to
$\Lambda_\ensuremath{{\chi\text{EFT} }}=600$ MeV and label the
results obtained by \ensuremath{{\chi\text{EFT} }}600.

In general, the effective low-momentum $YN$ interaction, the
$\ensuremath{V_\text{low k}}$ for hyperons, is obtained by solving a
renormalization group equation. The starting point for the
construction of the $\ensuremath{V_\text{low k}}$ is the half-on-shell
$T$-matrix. An effective low-momentum $\ensuremath{T_\text{low
    k}}$-matrix is then obtained from a non-relativistic
Lippmann-Schwinger equation in momentum space by introducing a
momentum cutoff $\Lambda$ in the kernel. Simultaneously, the bare
potential is replaced with the corresponding low-momentum potential
$\ensuremath{V_\text{low k}}$,
\begin{eqnarray}
  & &T_{\text{low }k, y' y}^{\alpha'\alpha}(q',q;q^2)=
    V_{\text{low }k,  y' y}^{\alpha'\alpha}(q',q)+ \nonumber \\
  &&\frac{2}{\pi}
  \sum_{\beta, z} P\!\!\int\limits_0^{\Lambda}\!\!dl\; l^2
  \frac{V_{\text{low }k, y' z}^{\alpha'\beta}(q',l)
        T_{\text{low }k, z
          y}^{\beta\alpha}(l,q;q^2)}{E_{y}(q)-E_{z}(l)}\ .
\label{eq:vlowk} 
\end{eqnarray}
The on-shell energy is denoted by $q^2$ while $q'$, $q$ are relative
momenta between a hyperon and nucleon. The labels $y$, $y'$, $z$
indicate the particle channels, and $\alpha$, $\alpha'$, $\beta$
denote the partial waves, e.g.~$\alpha=LSJ$ where $L$ is the angular momentum,
$J$ the total momentum and $S$ the spin. In Eq.~(\ref{eq:vlowk}) the energies
in the denominator are given by
\begin{eqnarray}
E_{y}(q)&=&M_{y}+\frac{q^2}{2\mu_{y}},
\label{eq:ene} 
\end{eqnarray}
with the reduced mass $\mu_{y}=M_Y M_N/M_y$, where $M_y$ is the total
mass  of the hyperon-nucleon system, $M_{y}=M_Y+M_N$. Finally,
the effective low-momentum $\ensuremath{V_\text{low k}}$ is defined by the
requirement that
the $T$-matrices are equivalent for all momenta below the cutoff
$\Lambda$. Details for several bare $YN$ potentials within the
RG framework can be found in Refs.~\cite{schaefer, wagner}.

\begin{figure}[h]
  \begin{center}
    \hspace{-1.5 cm}
    \includegraphics[width=10. cm]{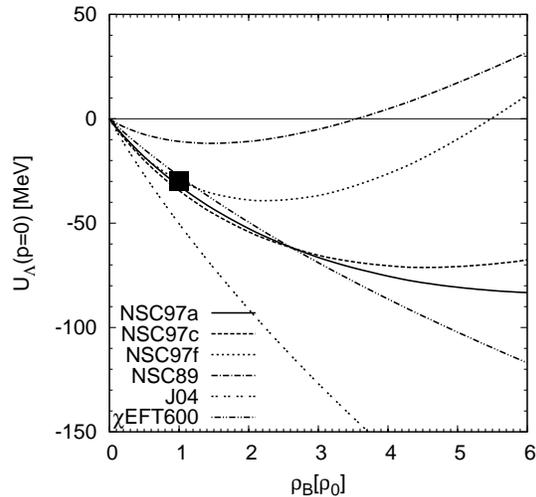}
    \vspace{-0.8 cm}
    \caption{Density dependence of $\Lambda$ single-particle
      potentials for various hyperon-nucleon interactions in symmetric
      nuclear matter.}
    \label{fig:Lrho}
  \end{center}
\end{figure}

From the different effective $\ensuremath{V_\text{low k}}$ interactions we
calculate the single-particle potential $U_{b}(p)$ of a baryon
$b \in \{ p,n, \Lambda, \Sigma^-, \Sigma^0, \Sigma^+, \Xi^-, \Xi^0 \}$ with
three-momentum $p = |\vec p |$. In general, it is defined as the
diagonal part in spin and flavor space of the proper self-energy for
the corresponding single-particle Green's function. In the
Hartree-Fock (HF) approximation for a uniform system it represents the
first-order interaction energy of the baryon with the filled Fermi
sea. It is evaluated as the diagonal elements of the low-momentum
potential matrix, $V_{y}^{\alpha} (q)$, where an evident short-hand
labeling for the diagonal elements has been introduced
(cf.~Eq.~(\ref{eq:vlowk})). Note, that the relative momentum is given
by $q=\left|\vec{p}-\vec{p}'\right|$ and the particle channel index is
given by $y=bb'$. In the HF approximation the single-particle
potential has two contributions: the (direct) Hartree- and the
(exchange) Fock-term~\cite{Walecka}
\begin{eqnarray}
  U_b(p)\!=\!\!\!\sum_{\vec{p}'\alpha b'}\!\!\left( 
    \left.V^{\alpha}_{y}(q)\right|_{\rm direct}\!\!+
    (-1)^{L+S} \left.V^{\alpha}_{y}(q)\right|_{\rm exchange}\right)\ .
\label{plane}
\end{eqnarray}
In this expression the diagonal elements of the nucleon-nucleon ($NN$)
interaction is also included. In principle, an effective low-momentum
potential for the $NN$ interaction is also known but we will use for
the $NN$-sector a parametric Ansatz to be discussed later (see
Eq.~(\ref{eq:ParEoS})).

\begin{figure}[htbp]
  \begin{center}
    \hspace{-1.5 cm}
    \includegraphics[width=10. cm]{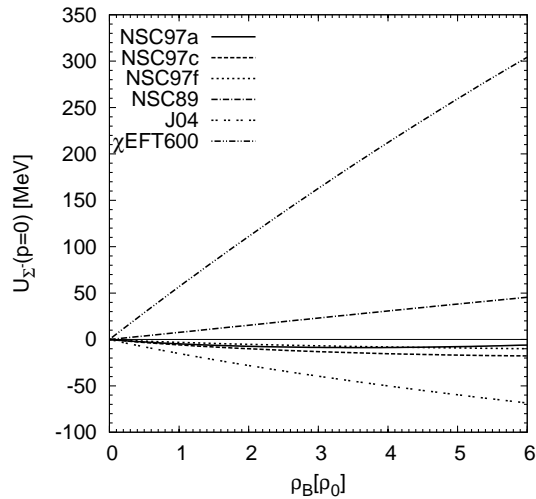}
    \vspace{-0.8 cm}
    \caption{Same as in Fig.~\ref{fig:Lrho} but for the $\Sigma^-$
      hyperon.}
    \label{fig:Smrho}
  \end{center}
\end{figure}

The density dependence for several $\Lambda$ potentials at rest in
symmetric nuclear matter (no hyperons present) is shown in
Fig.~\ref{fig:Lrho}. The square represents the generally excepted
empirical depth of $U_{\Lambda}(p=0)\approx -30\;\text{MeV}$. While
most of the potentials used can reproduce this value, the J\"ulich
potential (J04) yields a stronger binding while the old Nijmegen
potential (NSC89) underestimates the binding. All other potentials
agree up to the saturation density. However, with increasing density,
the differences  grow, leading to different
bindings at rest. This will have consequences for the predictions of
the $\Lambda$ hyperon concentration in dense nuclear matter. A
comparison with other works, \cite{schulze}, \cite{vidana1} and
\cite{rijken} shows some differences but mostly yields similar
results.

Fig.~\ref{fig:Smrho} shows the density dependence of several
$\Sigma^-$ potentials at rest in symmetric nuclear matter similar to
Fig.~\ref{fig:Lrho}. No agreement of the various $U_{\Sigma^-}$
potentials over the density range considered is seen. Compared with a
$G$-matrix calculation a stronger binding for the $\Sigma^-$
single-particle potential is obtained. In Ref.~\cite{Djapo:2008qv}
further details concerning the hyperon single-particle potentials can
be found.

For constructing the total energy/particle one has to distinguish
which part of the total single-particle potential comes from the
in-medium nucleon interaction and which from the hyperons. For
this purpose, the single-particle potential can be split into a
nucleonic and hyperonic contribution
\begin{eqnarray}
U_b(p)=U^N_b(p)+U^Y_b(p)\ ,
\label{eq:split}
\end{eqnarray}
where the first one, $U^N_{b}(p)$, denotes the baryon interacting with
a nucleon and the second one, $U^Y_{b}(p)$, means the baryon
interaction with a hyperon. Note, that the baryon can either be a
nucleon or a hyperon. Accordingly, the nucleonic contribution is
calculated from Eq.~(\ref{plane}) via
\begin{eqnarray}
  U^N_b(p)=\!\!\!\!\!\sum_{\begin{array}{c} \scriptstyle  \vec{p}' \alpha \\
      \scriptstyle b'={p,n} \end{array}}\!\!\!\!\!\left(
    \left.V^{\alpha}_{y}(q)\right|_{\rm direct}\!\!+ 
    (-1)^{L+S} \left.V^{\alpha}_{y}(q)\right|_{\rm exchange}\right)
\label{eq:Nb}
\end{eqnarray}
and analogously for the hyperonic contribution.

\section{Equation of State}
\label{sec:EoS}

By means of the single-particle potential the total energy per
particle, $E/A$, can be easily calculated. It is given by the
total energy of a baryon of mass $M_{b}$ and its kinetic
and potential energy $U_b (p)$ 
divided by the total baryon number density $\rho_B$:
\begin{eqnarray}
\label{E/A}
E/A&=&\frac{2}{\rho_B}\sum_{b}\int\limits_{0}^{k_{F_b}}\frac{d^3 p}{(2\pi)^3}
\left(M_{b}+\frac{p^2}{2M_{b}}+\frac{1}{2}U_{b}(p)\right).
\end{eqnarray}
The potential $U_{b}(p)$ describes the average field which acts on
these baryons due to their interaction with the medium. The baryon
Fermi momentum $k_{F_b}$ is given by
\begin{eqnarray}
  k^3_{F_b}=3\pi^2 x_b \rho_B
\end{eqnarray}
with the baryon fraction ratio $x_b=\rho_b/\rho_B$ for baryon $b$.

It is well-known that non-relativistic many-body calculations, based
on purely two-body forces, fail to reproduce the properties of nuclear
matter at saturation density. This is also the case in the present
work. In order to proceed we replace the purely nucleonic contributions
(without the influence of the hyperons) by an analytic parameterization
developed by Heiselberg and Hjort-Jensen \cite{Heiselberg:1999mq}.
This replacement makes the study of hyperons more robust since the
$NN$ sector can be controlled more easily.

From Eq.~(\ref{E/A}) the purely nucleonic contribution to the energy
per particle reads
\begin{eqnarray}
E_{NN}/A_N\!=\!\frac{2}{\rho_N}\!\sum_{N}\int\limits_{0}^{k_{F_N}}\!
\frac{d^3
p}{(2\pi)^3}\!\left(\!M_N+\frac{p^2}{2M_N}+\frac{1}{2}U^N_N(p)\!\right)
\label{nucleonic}
\end{eqnarray}
where the single-particle potential $U_N^N (p)$ only contains the
nucleonic contribution, cf.~Eq.~(\ref{eq:Nb}). However, instead of using 
the Hartree-Fock expression we employ the following parameterization
\begin{eqnarray}
E_{NN}/A_N =M_N-E_0u\frac{u-2-\delta}{1+u\delta}+S_0u^{\gamma}(1-2x_p)^2
\label{eq:ParEoS}
\end{eqnarray}
where $u=\rho_N/\rho_0$ denotes the ratio of the total nucleonic
density $\rho_N=(x_p+x_n)\rho_B$ to the nuclear saturation density
$\rho_0 =0.16\;\text{fm}^{-3}$. The corresponding proton and neutron
fraction are denoted by $x_p$ and $x_n$. The parameters
$E_0, \delta, S_0$ are related to properties of nuclear matter at
saturation density, i.e.~$E_0$ is the binding energy per nucleon at
saturation density while $S_0$ and $\delta$ are connected to the
symmetry energy and incompressibility, respectively.

As mentioned in the previous section, cf.~Eq.~(\ref{eq:split}), we can
separated the potential contribution of nucleons into one coming from
the interaction with other nucleons $U^N_N(p)$, and one coming from
the interaction with hyperons $U^Y_N(p)$. The latter does not
contribute to the purely nucleonic EoS when we consider symmetric
matter only. However, in the next section, when we investigate
hyperons such terms are considered.

The parameterization (\ref{eq:ParEoS}) is fitted to the energy per
particle in symmetric matter obtained from variational calculations
with the Argonne $V_{18}$ nucleon-nucleon interaction including
three-body forces and relativistic boost corrections
\cite{Akmal:1998cf}. The best fit parameters are
$E_0=-15.8\;\text{MeV}$, $S_0=32\;\text{MeV}$, $\gamma=0.6$ and
$\delta=0.2$. The EoS from Ref.~\cite{Akmal:1998cf} is considered as
one of the most reliable ones. In this way possible uncertainties
coming from the nucleonic EoS are minimized. 

The symmetry energy in dense matter is defined as
\begin{eqnarray}
  a_t=\left.\frac{1}{8} \frac{\partial^2  E/A}{\partial
      x_p^2}\right|_{\rho_B=\rho_0}\ .
\end{eqnarray}
Since there are no hyperons at saturation density we can use
Eq.~(\ref{eq:ParEoS}) directly to obtain $a_t=S_0$. The
incompressibility is given by
\begin{eqnarray} 
  K_0=\left.9\rho^2 \frac{\partial^2  E/A}{\partial
      \rho^2}\right|_{\rho_B=\rho_0} 
\end{eqnarray} 
from which we obtain the relation $K_0=-18E_0/(1+\delta)$.

\begin{figure}[!h]
  \begin{center}
    \includegraphics[width=8. cm]{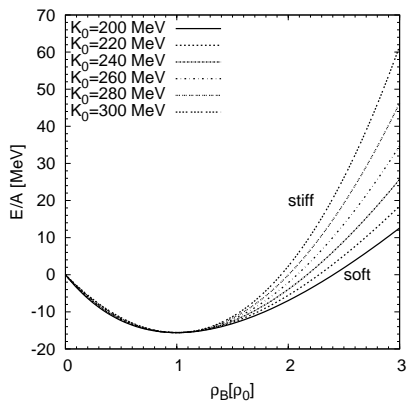}
    \vspace{-0.5 cm}
    \caption{Parametric EoS in symmetric nuclear matter as obtained
      from the parameterization of Ref.~\cite{Heiselberg:1999mq}.}
    \label{PEoS}
  \end{center}
\end{figure}

To study the impact of the softness of the EoS and the effects of the
symmetry energy we vary $K_0$ and $a_t$ in a broader interval.
From experimental constraints the values for $K_0$ range between
$200\;\text{MeV}$ and $300\;\text{MeV}$ and those for $a_t$ between
$28\;\text{MeV}$ and $36\;\text{MeV}$, see Ref.~\cite{Lattimer:2006xb}
and references therein. We vary $K_0$ and $a_t$ within these limits to
study the effects of both parameters on the appearance and
concentrations of hyperons in dense matter. While they do not
influence the particle concentrations directly they modify the
composition of the matter indirectly by changing the available energy. In
this way the point at which hyperons will appear is affected. Results
for various values of $K_0$ are shown in Fig.~\ref{PEoS} (in symmetric
matter the energy per particle is only sensitive to the
incompressibility). In all cases the saturation point is at $E/A=-16$
MeV. The parameter range allows us to classify the nucleonic EoS as a
stiffer ($K_0=300$ MeV) or a softer ($K_0=200$ MeV) one.

In addition, $K_0$ directly influences the maximum allowed mass of a
neutron star. By increasing $K_0$ the energy of the system is
increased and as a consequence, more and more hyperons can be
produced. This in turn will decrease the allowed maximum mass of a
neutron star. Such a nontrivial connection creates a conundrum: if we
use a stiffer nucleonic EoS by increasing $K_0$ we then allow for
higher hyperon concentrations which softens the total EoS.

By means of Eq.~(\ref{nucleonic}) we can split the total energy per
particle, Eq.~(\ref{E/A}), into a purely nucleonic part and a
remainder $E'/A$ via
\begin{eqnarray}
  E/A=\frac{\rho_N}{\rho_B}E_{NN}/A+E'/A
\label{E/A2}
\end{eqnarray}
with the rest
\begin{eqnarray}
  E'/A&=&\frac{2}{\rho_B}\sum_{N}\int\limits_{0}^{k_{F_N}}\frac{d^3 p}{(2\pi)^3}
  \frac{1}{2}U^Y_N(p)\nonumber\\&+&
  \frac{2}{\rho_B}\sum_{Y}\int\limits_{0}^{k_{F_Y}}\frac{d^3 p}{(2\pi)^3}
  \left(M_Y+\frac{p^2}{2M_Y}\right.\nonumber\\&+&\left.\frac{1}{2}U^N_Y(p)
    +\frac{1}{2}U^Y_Y(p)\right)\ .
\label{rest}
\end{eqnarray}

In symmetric matter, which is composed only of nucleons, $E'/A$
vanishes, but with this separation we can calculate the total $E/A$
for arbitrary hyperon concentrations. In the following we will
calculate the EoS including hyperons by determining their
concentrations in $\beta$-equilibrium.

\section{$\beta$-equilibrium}

\subsection{Composition of matter}
\label{sec:beta}

In order to determine the threshold densities for hyperons their
concentrations are needed. These are fixed by charge neutrality and
$\beta$-equilibrium. The latter refers to the equilibrium under the
weak interaction decays
\begin{eqnarray}
B_1\rightarrow B_2+l+\bar{\nu}_l
\end{eqnarray}
where $B_1$ and $B_2$ denote the baryons,
$l \in \{ e^-, \mu^-, \tau^- \}$ the negatively charged leptons and
$\bar{\nu}_l$ the corresponding neutrinos. In the case when the
neutrinos are not trapped in the star (i.e.~$\mu_{\nu}=0$) these
requirements amount to
\begin{eqnarray}
  \label{eq:q}
  0 &=&
  \sum_{b}(\rho^{(+)}_{b}-\rho^{(-)}_{b})+\sum_l(\rho^{(+)}_{l}-\rho^{(-)}_{l}) 
\end{eqnarray}
for the charge neutrality and 
\begin{eqnarray}
  \mu_{\Xi^-}=\mu_{\Sigma^-}&=&\mu_n+\mu_e,\label{eq:sne}\\
  \mu_{\Xi^0}=\mu_{\Lambda}=\mu_{\Sigma^0}&=&\mu_n,\label{eq:ln}\\
  \mu_{\Sigma^+}=\mu_p&=&\mu_n-\mu_e,\label{eq:sp}
\end{eqnarray}
for the chemical potentials. The densities of
positively and negatively charged baryons and leptons are denoted by
$\rho^{(\pm)}_b$ and $\rho^{(\pm)}_l$, respectively. The chemical
potentials $\mu$ are labeled by the corresponding particles. In the
absence of neutrinos all lepton and antilepton chemical potentials
are equal. In addition to the electrons, muons are also present. The 
$\tau$-lepton does not appear since it is too heavy.

\begin{figure}[h]
  \begin{center}
    \hspace{-1.1cm}
    \includegraphics[width=9 cm]{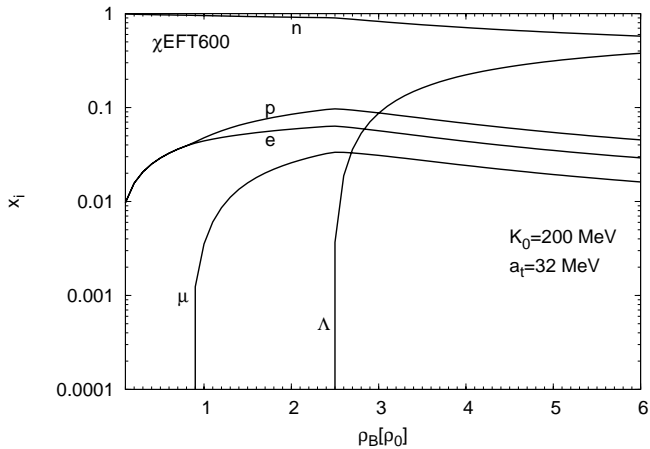}
    \caption{Density ratios for different particles for a "soft" 
    nucleonic EoS
      as a function of the baryon density using the 
    \ensuremath{{\chi\text{EFT}}}600 model.}
    \label{200.32.60}
  \end{center}
\end{figure}

At zero temperature, the chemical potential of a fermion system is equal to
its Fermi energy. For relativistic noninteracting leptons it is given by
\begin{eqnarray}
  \mu_l=\sqrt{m_l^2 + k_{F_l}^2}= \sqrt{m_l^2+ \left(3\pi^2
      x_l\rho\right)^\frac{2}{3}}\ ,
\end{eqnarray}
with the corresponding lepton density ratio $x_l = \rho_l/\rho_L$. The
total lepton density $\rho_L$ is the sum over all three leptons. For
nonrelativistic interacting baryons, the chemical potential for
species $b$ reads
\begin{eqnarray}
  \mu_{b}=M_{b}+\frac{k_{F_b}^2}{2M_{b}}+U_{b}(k_{F_b})\ .
\label{eq:munr}
\end{eqnarray}

For a given total baryon density $\rho_B$ the equations
(\ref{eq:q})-(\ref{eq:sp}) govern the composition of matter, i.e. the
baryonic and leptonic concentrations. The corresponding solution is
referred to as $\beta$-stable matter.

For the sake of consistency we now have to treat the nucleonic part of
the chemical potential $\mu_N$ in the same way as the corresponding
energy per particle. Since the chemical potential can be obtained as a
derivative of the energy density $\epsilon$ and is related to the
energy per particle via $\epsilon=\rho_B E/A$, we use the definition
\begin{eqnarray} 
  \mu_b=\frac{\partial \epsilon}{\partial \rho_b}\ ,
\end{eqnarray}
to have the appropriate replacement in the nucleonic chemical
potential. Finally, we arrive at the expression
\begin{eqnarray}
  \mu_N=\frac{\partial \epsilon_{NN}}{\partial \rho_N}+U^Y_N(k_{F_Y}),
  \label{eq:mupara}
\end{eqnarray}
where we have effectively replaced
$M_N+\frac{k^2_{F_N}}{2M_N}+U^N_N(k_{F_N})$ of Eq.~(\ref{eq:munr})
with the derivative $\partial \epsilon_{NN}/\partial \rho_N$. In this
way the parameterization Eq.~(\ref{eq:ParEoS}) enters in the nucleonic
part of the chemical potential.

\begin{figure}[h]
  \begin{center}
    \hspace{-1.1cm}
    \includegraphics[width=9 cm]{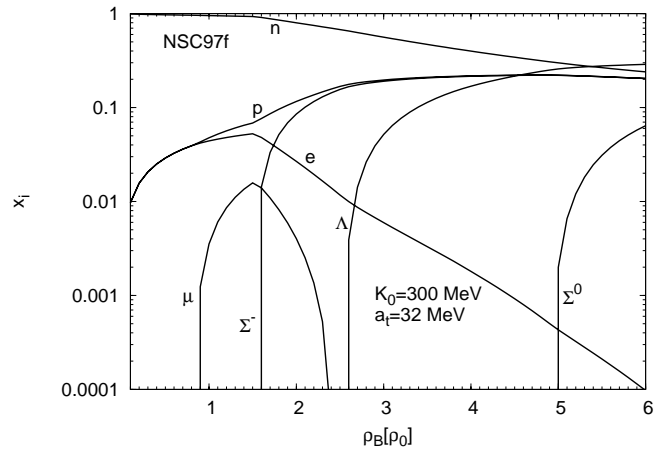}
    \caption{Same as Fig.~\ref{200.32.60} but for "stiff" nucleonic EoS using
      the NSC97f model.}
    \label{300.32.Nf}
  \end{center}
\end{figure}

Since we are only parameterizing the nucleonic sector, no such
replacement is necessary for the hyperons. However, since we have
neglected the $YY$ interaction $U^Y_Y(k_{F_Y})$ is zero and
Eq.~(\ref{eq:munr}) reduces to
\begin{eqnarray}
  \mu_Y=M_Y+\frac{k^2_{F_Y}}{2M_Y}+U^N_Y(k_{F_Y})\ .
\end{eqnarray}

\begin{figure}[h]
  \begin{center}
\includegraphics[width=9. cm]{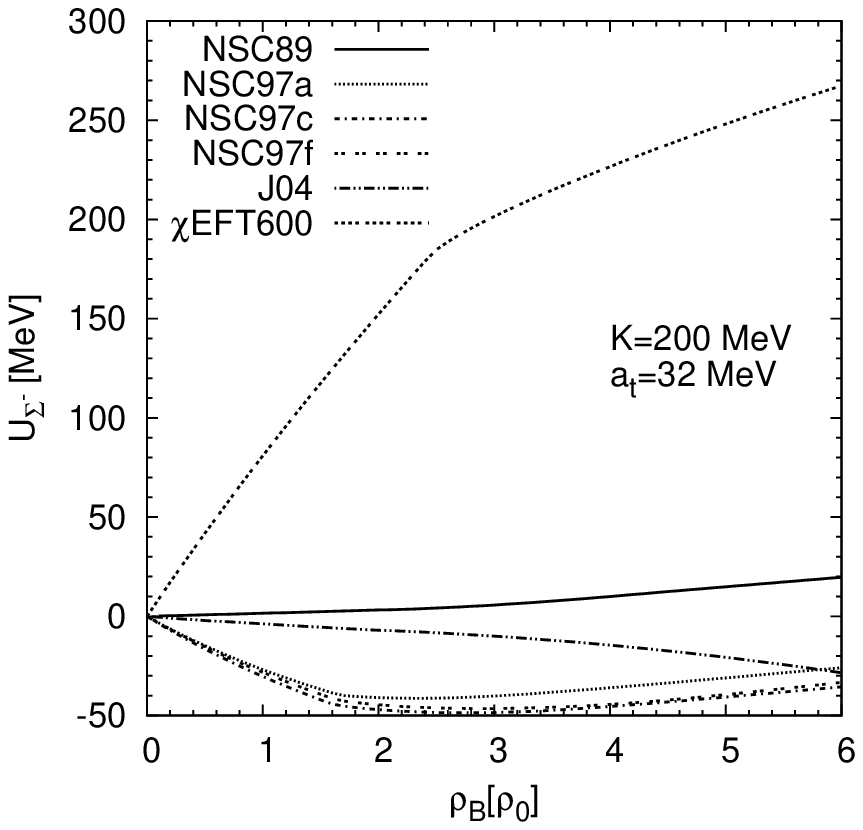}\\ 
\vspace{-0.5 cm}
\includegraphics[width=9. cm]{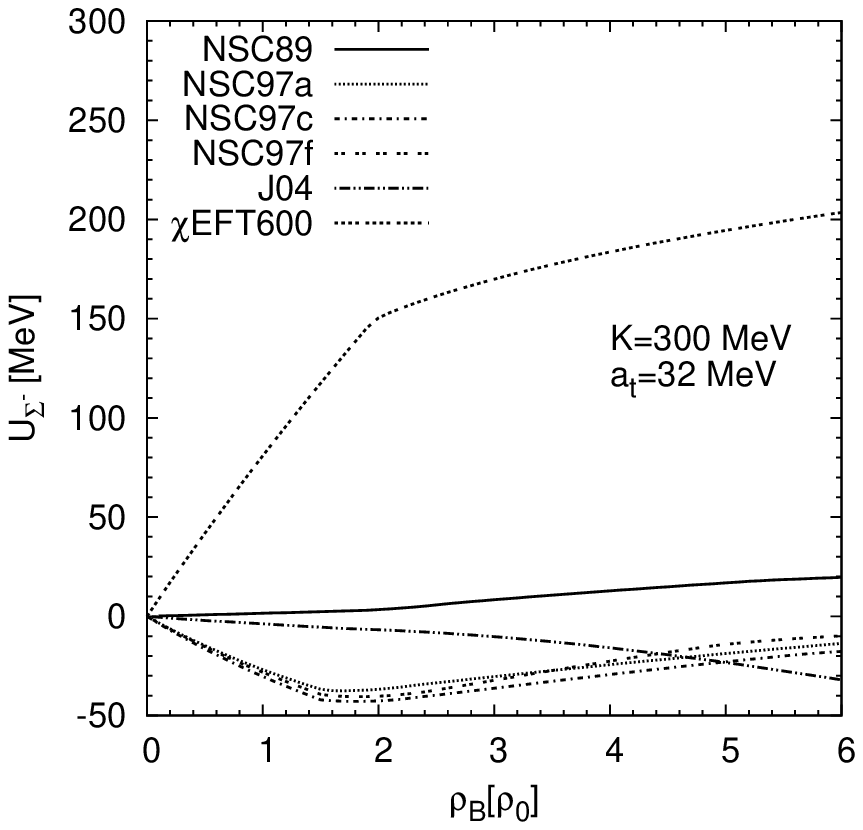}
\vspace{-1. cm}
\caption{Density dependence of $U_{\Sigma^-} (k_{F_{\Sigma^-}})$ for
  $\beta$-equilibrated matter. Upper panel: $K_0=200$ MeV, lower
  panel: $K_0=300$ MeV.}
\label{fig:usmpf}
\end{center}
\end{figure}

For the determination of the particle concentration the
single-particle potential in equilibrium is used. For hyperons below
the threshold density it is given by $U_Y(p=0)$, similar to the
symmetric matter case. Above the threshold density it depends on
composition and density. In Fig.~\ref{fig:usmpf} the density
dependence of $U_{\Sigma^-}(k_{F_{\Sigma^-}})$ in $\beta$-equilibrium
for two different incompressibilities $K_0$ is shown. In the figure a
kink in the curves appears at the point where the hyperons appear.
 
Another observation is the relative ordering and the magnitudes which
resemble those of the single-particle potentials at zero momentum in
symmetric matter as shown in Fig.~\ref{fig:Lrho}. Essentially, the
NSC97a, NCS97c, NSC97f and J04 interactions are still slightly
attractive while the NSC89 and $\chi$EFT600 remain repulsive. Similar
observations hold for the $\Lambda$ system. A new structure in form of
a second inflection point emerges due to the appearance of the
$\Sigma^-$ hyperon.

A better indicator at which densities hyperons start to appear is
given by the concentrations of all particles and is displayed in
Figs.~\ref{200.32.60} and \ref{300.32.Nf}. In Fig.~\ref{200.32.60} a
``soft'' nucleonic EoS is used in combination with an attractive
$\Lambda N$ and a very repulsive $\Sigma N$ interaction implemented by
the \ensuremath{{\chi\text{EFT} }}600 model. In contrast in
Fig.~\ref{300.32.Nf} a ``stiff'' EoS is used represented by the NSC97f
model which has a similar $\Lambda N$ interaction compared to the
\ensuremath{{\chi\text{EFT} }}600 model but also an attractive $\Sigma
N$ interaction. This difference already leads to very different
density profiles. While in Fig.~\ref{200.32.60} the $\Lambda$ hyperon
is the first one which appears and no $\Sigma^-$ hyperons are present,
the $\Sigma^-$ hyperon appears first in Fig.~\ref{300.32.Nf}.

One should note that with the appearance of the $\Sigma^-$ hyperon the
density of the negatively charged leptons starts to drop immediately.
This is because their role in the charge neutrality condition,
Eq.~(\ref{eq:q}), is now being taken over by the $\Sigma^-$.
Similarly, the appearance of the $\Lambda$ hyperon will accelerate the
disappearance of neutrons since both are neutral particles.

\begin{figure}[h]
  \begin{center}
\includegraphics[width=9. cm]{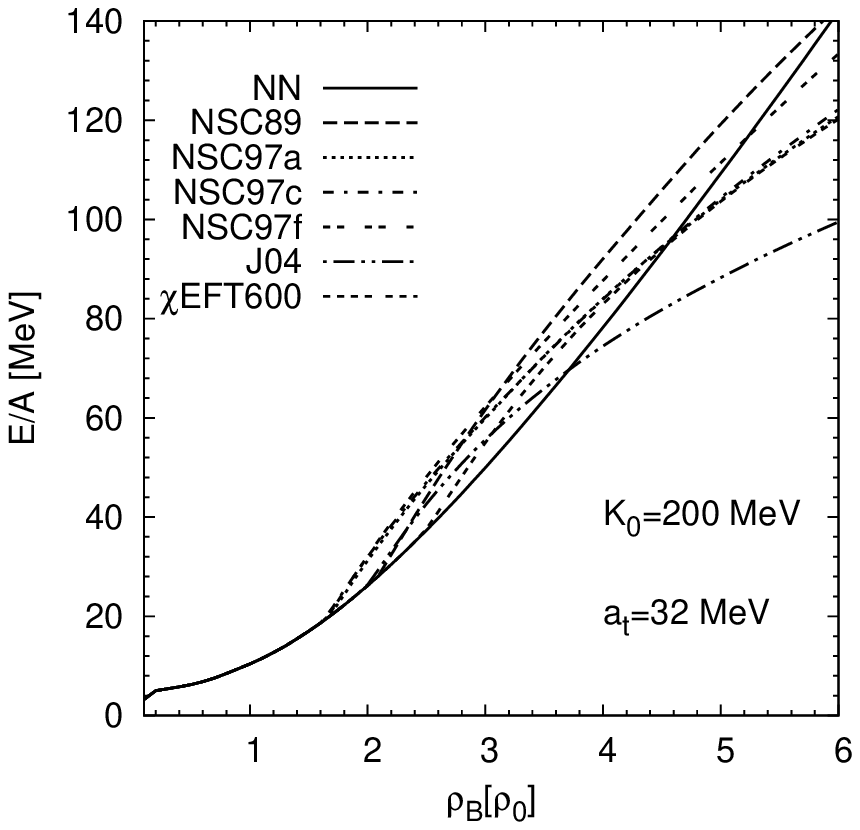}\\ 
\vspace{-0.5 cm}
\includegraphics[width=9. cm]{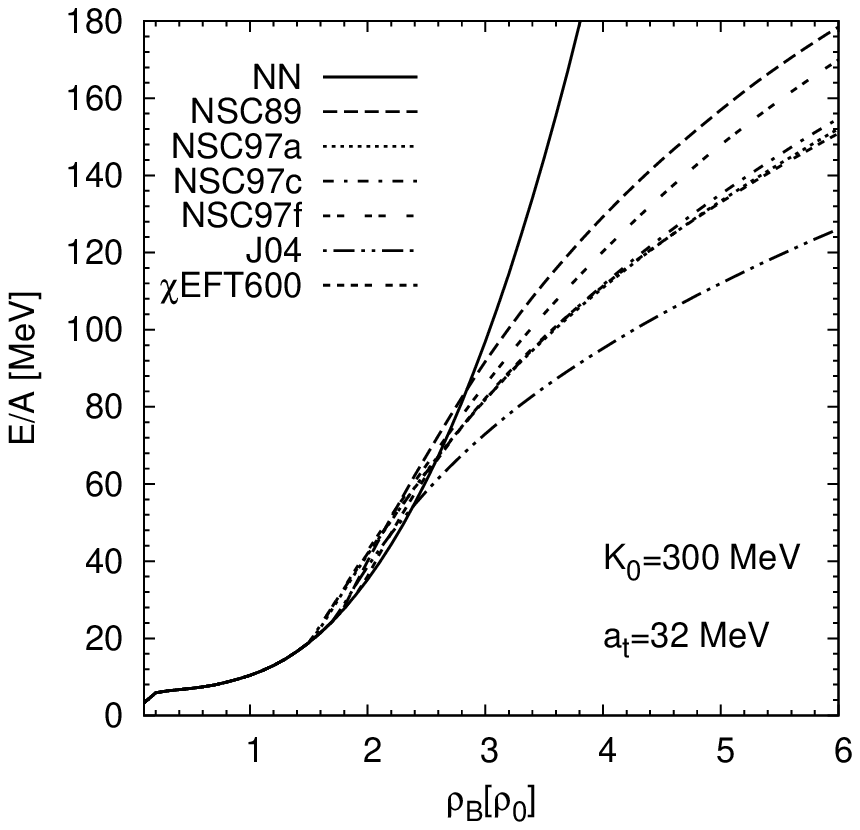}
\vspace{-1. cm}
\caption{Various EoS for $\beta$-equilibrated matter for two different
  $K_0$ as a function of density. Upper panel: soft EoS,
  lower panel: stiff EoS.}
\label{bEoS}
\end{center}
\end{figure}

Once the composition of the matter has been determined by demanding
$\beta$-equilibrium we can calculate the energy per particle. For this
purpose, we cannot use Eq.~(\ref{E/A}), but have to use
Eqs.~(\ref{E/A2}) and (\ref{eq:ParEoS}). The result is presented in
Fig.~\ref{bEoS} where the energy per particle in $\beta$-stable matter
is shown as a function of the density for different $YN$ models. The
symmetry energy is fixed to $a_t = 32$ MeV while the incompressibility
is set to $K_0=200$ MeV (upper panel in the figure) and to $K_0=300$
MeV (lower panel). In addition, the EoS with hyperons is compared with
the purely nucleonic one.

One easily observes the onset of the hyperon appearance at the point
at which the curves start to deviate. As expected the differences
between the various $YN$ interactions do not modify the EoS for very
small densities. In the range between $(2 - 3)\rho_0$, all EoS's are
similar to each other. However, for increasing densities the influence
of hyperons becomes more significant resulting in rather different
EoS's. This concerns not only the magnitudes of the different energies
per particle but also their slopes at higher densities. These
variations will lead to differences in the pressure and finally to
significant changes in the possible maximum mass of a neutron star.

\subsection{Threshold densities}
\label{sec:Threshold densities}

\begin{figure*}[ht]
  \begin{center}
      \vspace{-6.cm}
      \includegraphics[width=18. cm]{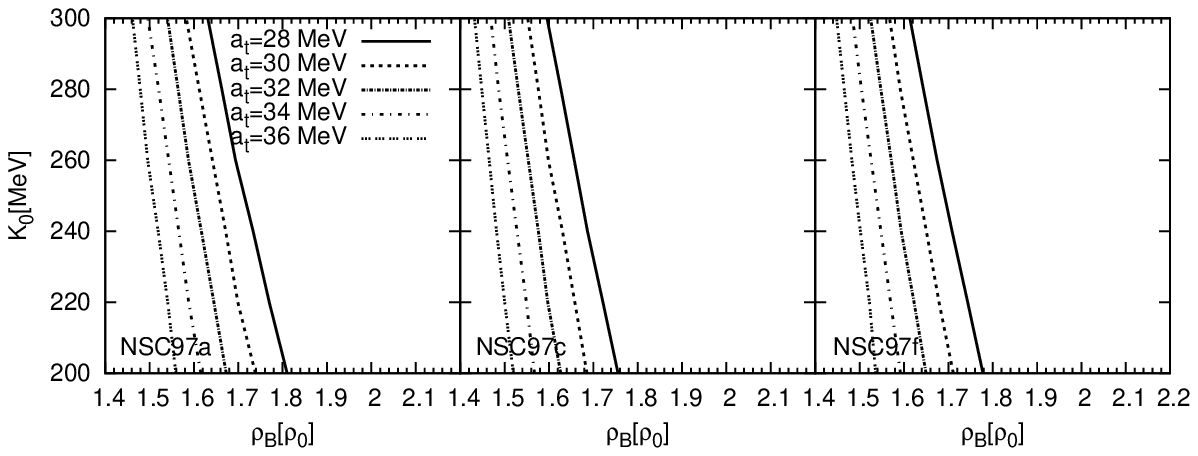}\\
      \vspace{-6.cm}      
      \includegraphics[width=18. cm]{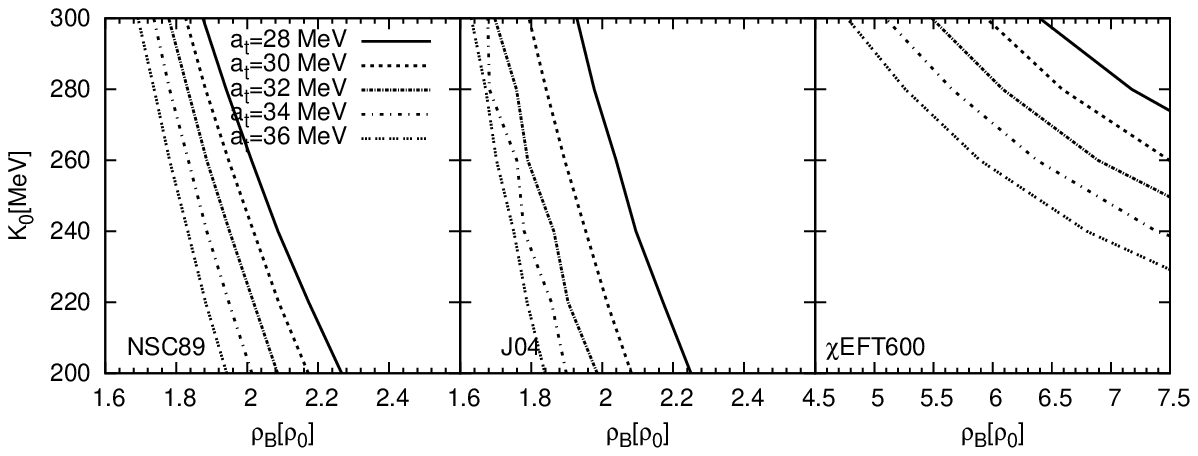}
      \vspace{-1.cm}    
      \caption{Threshold densities for the $\Sigma^-$ hyperon with
        various incompressibilities $K_0$ and symmetry energy
        $a_t$ for six different YN interactions.}
     \label{fig:stre}
  \end{center}
\end{figure*}

\begin{figure*}[ht]
  \begin{center}
      \vspace{-6.cm}
      \includegraphics[width=18. cm]{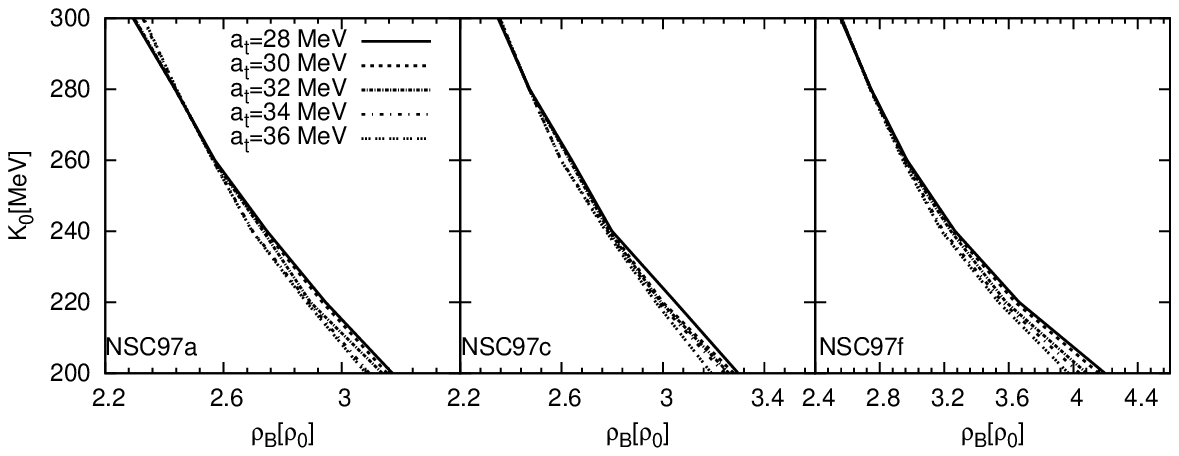}\\
      \vspace{-6.cm}      
      \includegraphics[width=18. cm]{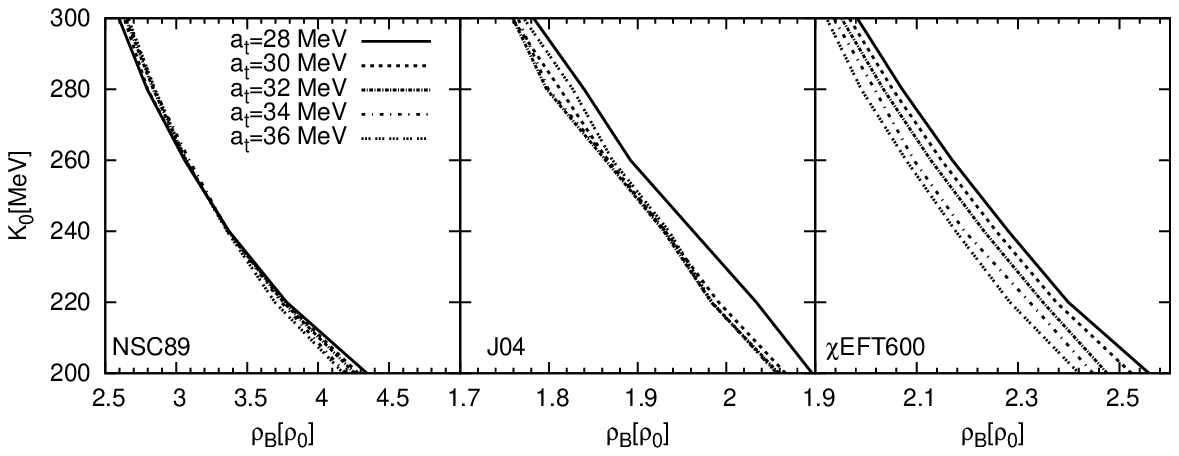}
      \vspace{-1.cm} 
      \caption{Threshold densities for the $\Lambda$ hyperon similar
        to Fig.~\ref{fig:stre}.}
     \label{ltre}
  \end{center}
\end{figure*}

The appearance of a given hyperon species is determined by increasing
the density for fixed $K_0$ and $a_t$. The resulting threshold
densities for the $\Sigma^-$ hyperon for certain $K_0$ and $a_t$ are
collected in Fig.~\ref{fig:stre} for six different $YN$ interactions.
Similarly, the threshold densities for the $\Lambda$ hyperon are shown
in Fig.~\ref{ltre}. From these figures one sees how the single-particle potentials for
various $YN$ interactions modify the threshold densities. In this way,
the properties of the $YN$ interaction in Fig.~\ref{fig:Lrho} and
Fig.~\ref{fig:Smrho} can be attributed to the hyperon appearances.

From Fig.~\ref{fig:stre} one concludes that the $\Sigma^-$ hyperon appears
between $1.4 \rho_0$ and $2.4\rho_0$ with the exception of the
\ensuremath{{\chi\text{EFT} }}600 model. For almost all $YN$
interactions used in the present study the $\Sigma^-$ is the first
hyperon which will appear even though the $\Lambda$ hyperon is the
lighter one. The reason is that the heavier mass of
the $\Sigma^-$ is offset due to the presence of the $e^-$ chemical
potential, cf.~Eq.~(\ref{eq:sne}). In general, heavier and more
positively charged particles appear later. In the case of the
$\Sigma^-$, compared to the $\Lambda$, the effect caused by the
electric charge dominates the one coming from the mass in almost all
cases.
For the $\Sigma^-$ hyperon a further modification caused by the
electric charge, is the influence of  $a_t$ on the threshold
density because the electron chemical potential is modified by the
symmetry energy. Thus, the decrease of the threshold densities due to
the increase of $K_0$ is analogous to the increase due to $a_t$.

For the $\Lambda$ hyperon the range of threshold densities is
between densities from $1.7 \rho_0$ to $4.5 \rho_0$ depending on the
choice of $K_0$, $a_t$ and the $YN$ interaction used,
cf.~Fig.~\ref{ltre}. The influence of $K_0$ on the
threshold density for this hyperon is larger than the one from 
$a_t$. This is reasonable since $K_0$
controls the rate of the energy increase with the density more
directly, while $a_t$ affects only the details of the
$\beta$-equilibrium. One clearly recognizes in Fig.~\ref{ltre} that the
$\Lambda$ appears earlier for larger incompressibilities. Thus, in
general we see that for increasing $K_0$ the threshold
densities decrease for both hyperons.

In contrast to the previous $K_0$ and $a_t$ discussion, the influence
of the single-particle potentials on the threshold densities is harder
to analyze.

The threshold densities for the $\Sigma^-$ are largest for the
\ensuremath{{\chi\text{EFT} }}600 interaction which yields the most
repulsive $\Sigma^-$ single-particle potential. In general, hyperons
will appear earlier for a more attractive single-particle potential.
This becomes obvious from Eq.~(\ref{eq:munr}): the chemical potential
decreases for a more negative $U_b(k_{F_b})$ and, consequently, the
threshold density will also decrease. Thus, the most repulsive
single-particle potential like the one for the
\ensuremath{{\chi\text{EFT} }}600 leads to the largest threshold
density. For the $\Lambda$ hyperon the threshold densities are
smallest for the most attractive single-particle potential obtained
with the J04 model, cf.~Fig.~\ref{ltre}. On the other hand, they are
largest for the most repulsive NSC89 interaction. For the NSC97f
interaction, which is between these extremes, the $\Lambda$ threshold
densities are very close to those of the most repulsive NSC89 one,
cf.~Fig.~\ref{ltre}. This stems from the appearance of the $\Sigma^-$
hyperon. The effect is caused by the slowdown of the increase of the
neutron chemical potential and is further related to the rapid
increase of the $\Sigma^-$ density just after its appearance,
cf.~Fig.~\ref{300.32.Nf}. Basically, the slowdown occurs as soon as a
new hyperon appears because most of the energy is used for its
creation. Once the concentration of the hyperon has reached a plateau,
the neutron chemical potential resumes its increase until a further
hyperon might appear. Thus, the appearance of the first hyperon shifts
the threshold density of the next hyperon towards higher values. This
effect explains why the threshold densities of the $\Lambda$ are so
similar for the NSC97f and NSC89 interactions. It also makes clear why
the $\Lambda$ threshold densities for the \ensuremath{{\chi\text{EFT}
  }}600 interaction are smaller than those of the NSC97a, NSC97c and
NSC97f interactions even though their $\Lambda$ single-particle
potentials are almost the same, cf.~Fig.~\ref{fig:Lrho}. In the case
of the J04 model the delay mechanism described above becomes very
interesting. For this $YN$ interaction the $\Lambda$ and $\Sigma^-$
hyperon appear almost at the same density. In this case the neutron
chemical potential stagnates but the $\Lambda$ and the $\Sigma^-$
single-particle potentials are attractive enough to compensate for
this.

To summarize this section strangeness appears around $\sim 2\rho_0$
for all $YN$ models and parameter sets used. Note, that the appearance
of the first hyperon, whether it is the $\Sigma^-$ or the $\Lambda$,
cannot be altered by taking into account $YY$ interactions which have
been neglected in this work. The present study in terms of the broad
parameter ranges as well as the multitude of $YN$ interaction models
reveals that strangeness in the interior of neutron stars cannot be
ignored. Similar conclusions are obtained in the
Brueckner-Hartree-Fock theory \cite{Baldo:1998hd}.

\section{Structure of neutron stars}
\label{sec:structure}

In this section we analyze the effect of the EoS including hyperons on
neutron stars. We focus on non-rotating stars, ignoring any changes,
caused by the rotation. For a given EoS, the mass-radius relation of a
NS can be determined by solving the familiar
Tolman-Oppenheimer-Volkoff equation (TOV) \cite{Oppenheimer:1939ne}.
To describe the outer crust and atmosphere of the star i.e., the
region of subnuclear matter densities for very small baryon densities
below $\rho_B < 0.001\;\textrm{fm}^{-3}$, we have used the EoS of
Baym, Pethick, and Sutherland \cite{Baym:1971pw}, which relies on
properties of heavy nuclei. For densities between
$0.001\;\textrm{fm}^{-3} \leq \rho_B \leq 0.08\;\textrm{fm}^{-3}$,i.e.
for the inner crust, we have used the EoS of Negele and Vautherin
\cite{Negele:1971vb}, who have performed Hartree-Fock calculations of
the nuclear crust composition. Details on crust properties can be
found e.g.~in \cite{Pethick:1995di,Blaschke:2001uj}, while recent
state-of-the-art approaches are discussed in \cite{Ruester:2005fm}.

\begin{figure*}[ht]
  \subfigure[$\ K_0=200\;\text{MeV}$]{\label{fig:mR200}
 \hspace{-0.5 cm}\includegraphics[width=0.58\textwidth]{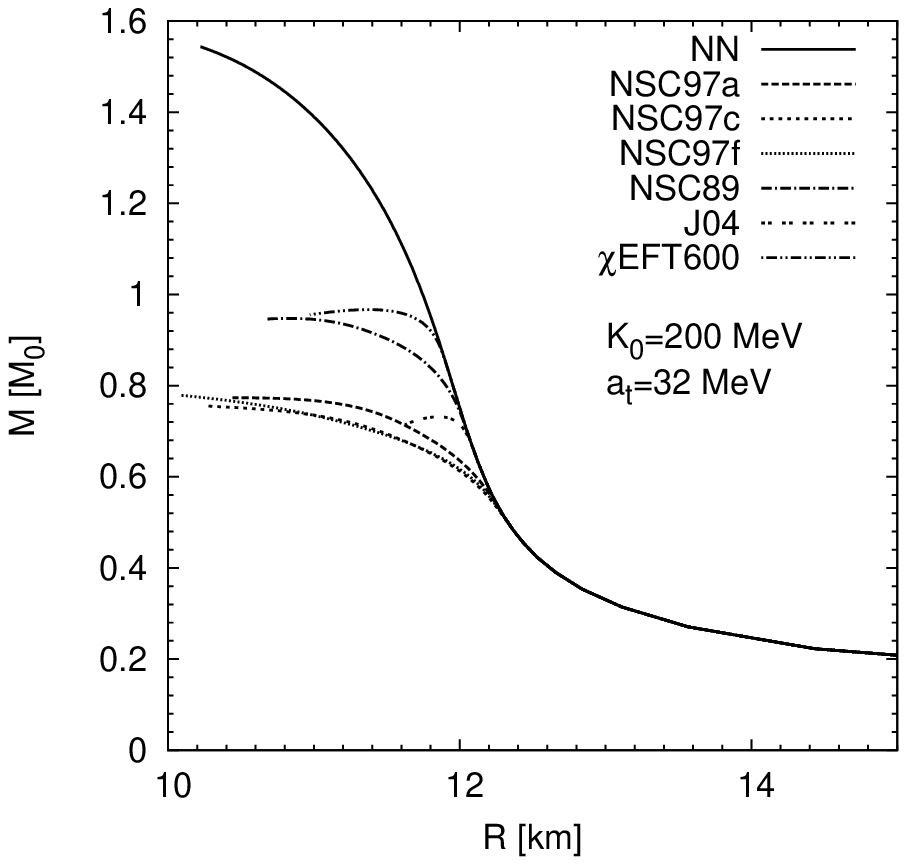}}\hspace{-3. cm}
  \subfigure[$\ K_0=300\;\text{MeV}$ ]{\label{fig:mR300}
    \includegraphics[width=0.58\textwidth]{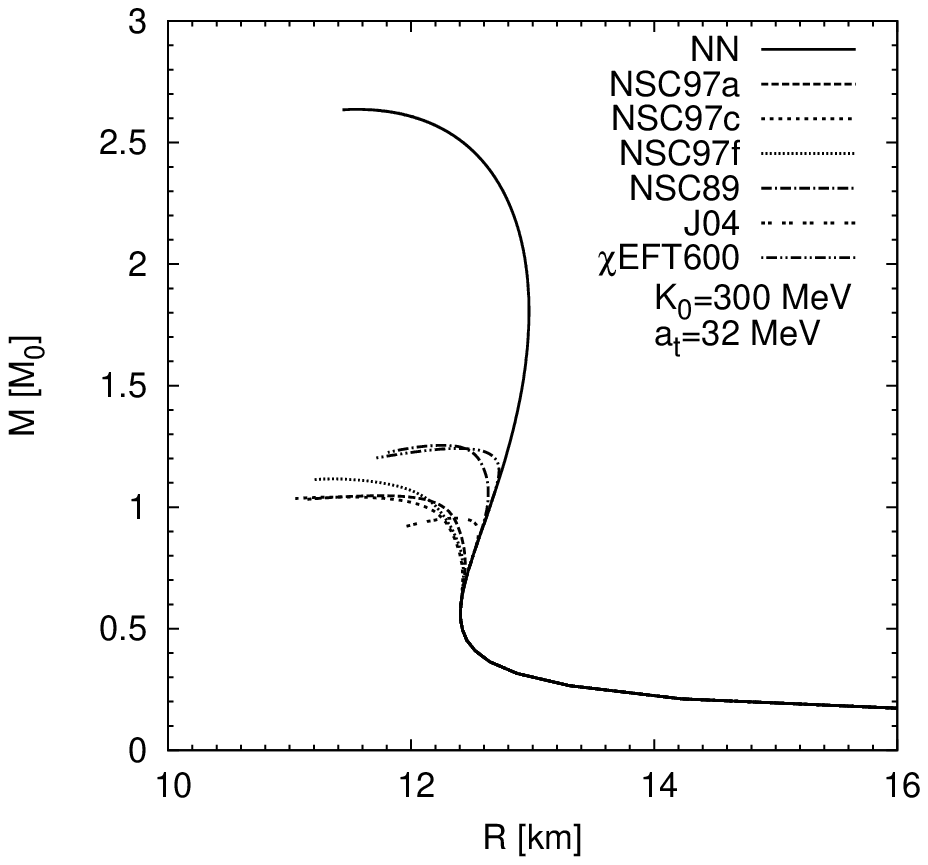}}\hspace{-0.5 cm}
  \caption{Mass-radius relation of a neutron star for symmetry energy
    $a_t=32\;\text{MeV}$ and different $YN$ interactions. For
    comparison the mass-radius curve obtained for the pure $NN$
    interaction is also shown. Left panel: soft EoS, right panel:
    stiff EoS. \label{fig:mR}}
\end{figure*}

In Fig.~\ref{fig:mR} the mass-radius relation of a NS for a soft EoS (left
panel) and for a stiff EoS (right panel) is shown. The symmetry
energy $a_t = 32$ MeV is kept fixed in both calculations and the
resulting mass-radius relation without any strangeness is
also added for comparison.

As can be seen from the figure the appearance of hyperons reduces the
NS mass drastically compared with the pure $NN$ case. Even for larger
values of the incompressibility, i.e.~$K_0=300$ MeV, the maximum mass,
obtained with all $YN$ interactions used, is still below the largest
precisely known and measured NS masses $1.44\;M_{\odot}$ of the
Hulse-Taylor binary pulsar. This is not an unusual result and is also
seen in other, related works such as e.g.~\cite{Vidana:2000ew,
  SchaffnerBielich:2002ki, SchaffnerBielich:2007yq, Mornas:2004vs}. In
general, any inclusion of further degrees of freedom to the nucleons
will reduce the NS mass.

\begin{figure*}[ht]
  \subfigure[$\ K_0=200\;\text{MeV}$,$\gamma=1 \text{ for } \rho\geq\rho_0$]{\label{fig:m2R200}
 \hspace{-0.5 cm}\includegraphics[width=0.58\textwidth]{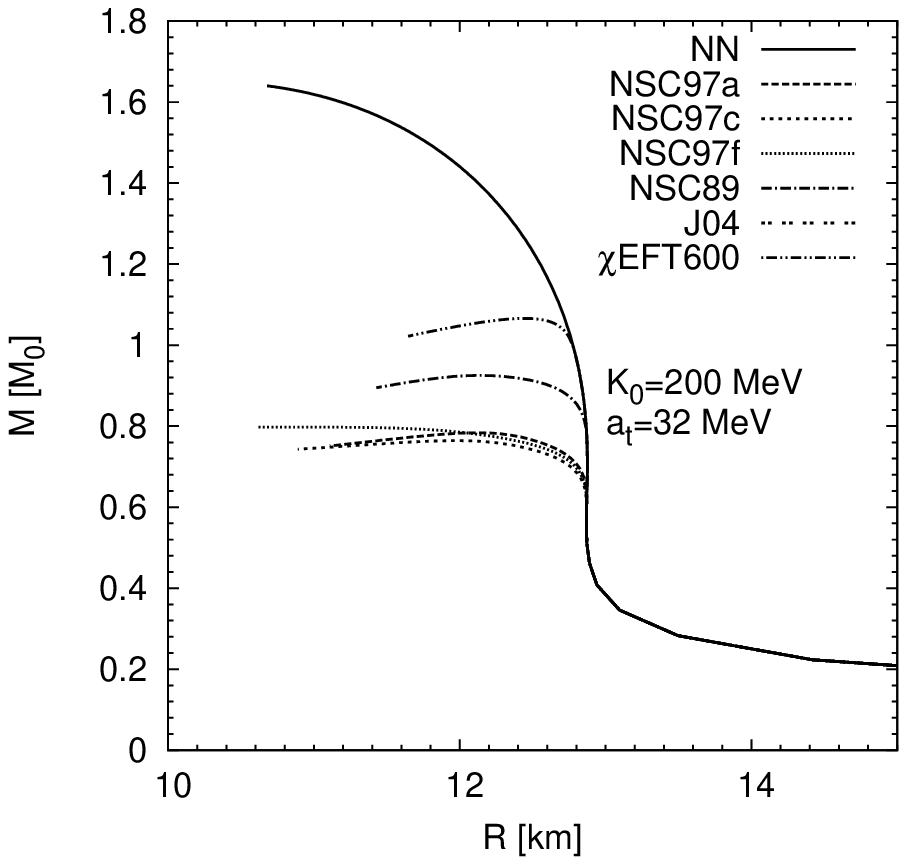}}\hspace{-3. cm}
  \subfigure[$\ K_0=300\;\text{MeV}$,$\gamma=1 \text{ for } \rho\geq\rho_0$]{\label{fig:m2R300}
    \includegraphics[width=0.58\textwidth]{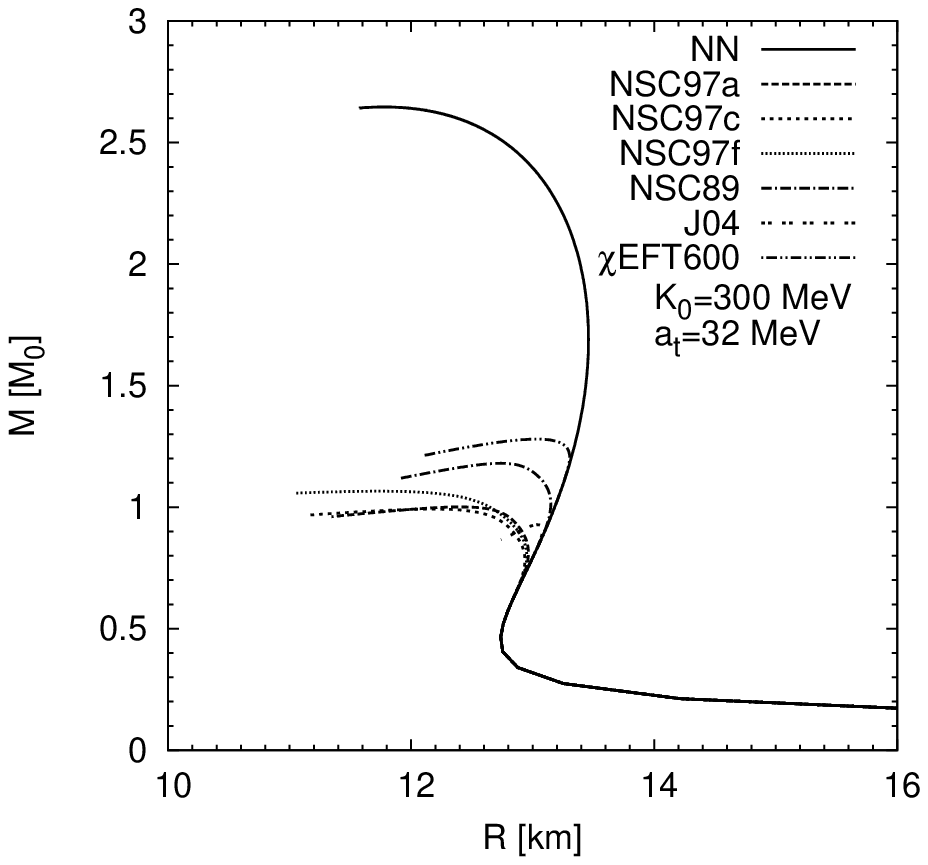}}\hspace{-0.5 cm}
  \caption{Mass-radius relation of a neutron star for symmetry energy
    $a_t=32\;\text{MeV}$, $\gamma=1 \text{ for } \rho\geq\rho_0$ and different $YN$ interactions. For
    comparison the mass-radius curve obtained for the pure $NN$
    interaction is also shown. Left panel: soft EoS, right panel:
    stiff EoS. \label{fig:m2R}}
\end{figure*}

Due to large uncertainties in the high density behavior of the
symmetry energy we investigate its influence on the NS masses as
follows: The parameter $\gamma$ in Eq.~(\ref{eq:ParEoS}) determines
the symmetry energy changes, i.e., how asy-stiff or asy-soft the EoS
is. The value of $\gamma = 0.6$, used so far, represents a asy-soft
system \cite{Wolter:2008zw}. Above saturation density we change the
value of this parameter and use $\gamma=1$ while keeping all other
parameters fixed representing a asy-stiff system. The effect of the
$\gamma$ modification is shown in Fig.~\ref{fig:m2R}.

Only small changes in the mass-radius relation are obtained when
hyperons are included. In most cases the mass difference is $~0.1
M_\odot$. An increasing $\gamma$ leads to an increase in the proton
concentration which in turn leads to an increase in the $\Sigma^-$
concentration. This largely cancels any energy gain from an increased
symmetry energy term that could possibly increase the mass of a
neutron star. The biggest mass change is seen for $K_0=200$ MeV with
the $\chi$EFT600 interaction since this model does not contain any
$\Sigma^-$ hyperons. However, even in this case the difference is
below $0.2 M_\odot$ which leads to a maximum mass below $1.2 M_\odot$.
A more noticeable change can be observed for the radius. A radius
shift of 1 km towards larger radii is seen for all curves.
Furthermore, if we in addition vary $a_t$ at saturation density
between 28 and 36 MeV, a less pronounced effect on the NS mass is
visible as compared to the variation of $\gamma$.

To complete our investigation of hyperon effects on the neutron star
EoS we also need to consider modifications induced by the $YY$ and
$\Xi N$ interactions. In order to evaluate the effects of the $S=-2$
sector we have to construct $YY$ and $\Xi N$ $\ensuremath{V_\text{low
    k}}$ interactions, along the lines discussed earlier for the
$S=-1$ sector. Obviously, since there are more particles to consider,
the situation complicates considerably from a numerical point of view.
Unfortunately, unlike the $S=-1$ sector where we had several
different ``bare'' interactions from which we constructed the
$\ensuremath{V_\text{low k}}$ potentials, in the $S=-2$ sector we only
have one, namely the NSC97 interaction. We also note that the
inclusion of the $S=-2$ sector will not influence the appearance of
the first hyperon. Hence we have neglected the $S=-2$ up to this
point.

\begin{figure*}[ht]
  \subfigure[$\ K_0=200\;\text{MeV}$]{\label{fig:mYYR200}
 \hspace{-0.5 cm}\includegraphics[width=0.58\textwidth]{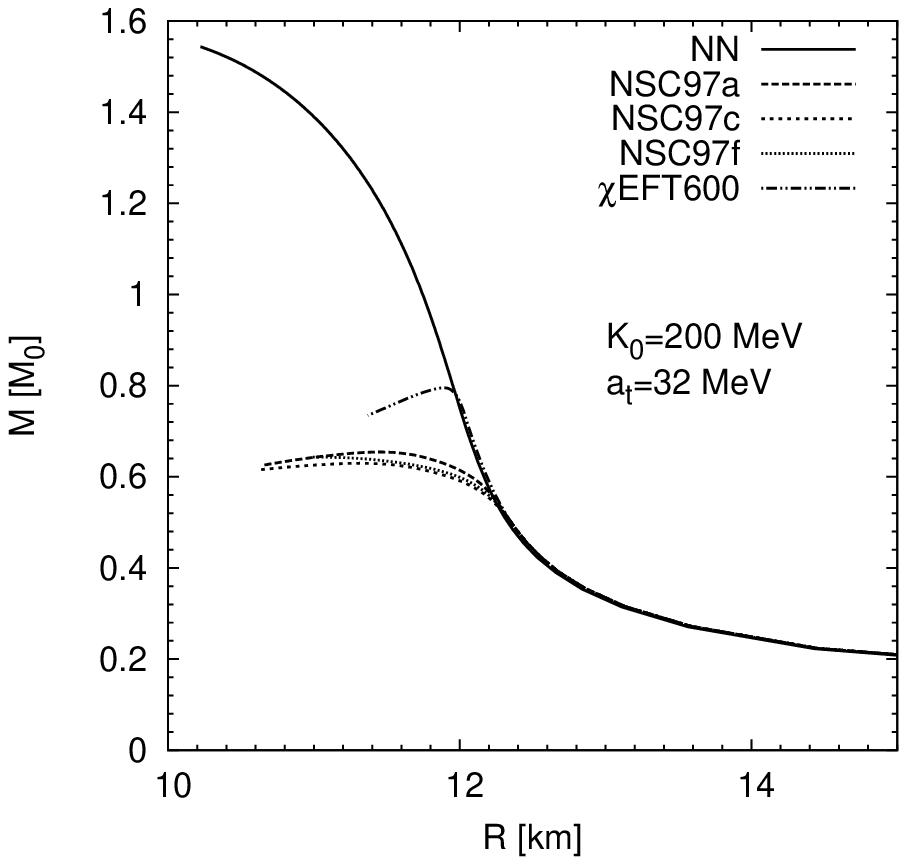}}\hspace{-3. cm}
  \subfigure[$\ K_0=300\;\text{MeV}$]{\label{fig:mYYR300}
    \includegraphics[width=0.58\textwidth]{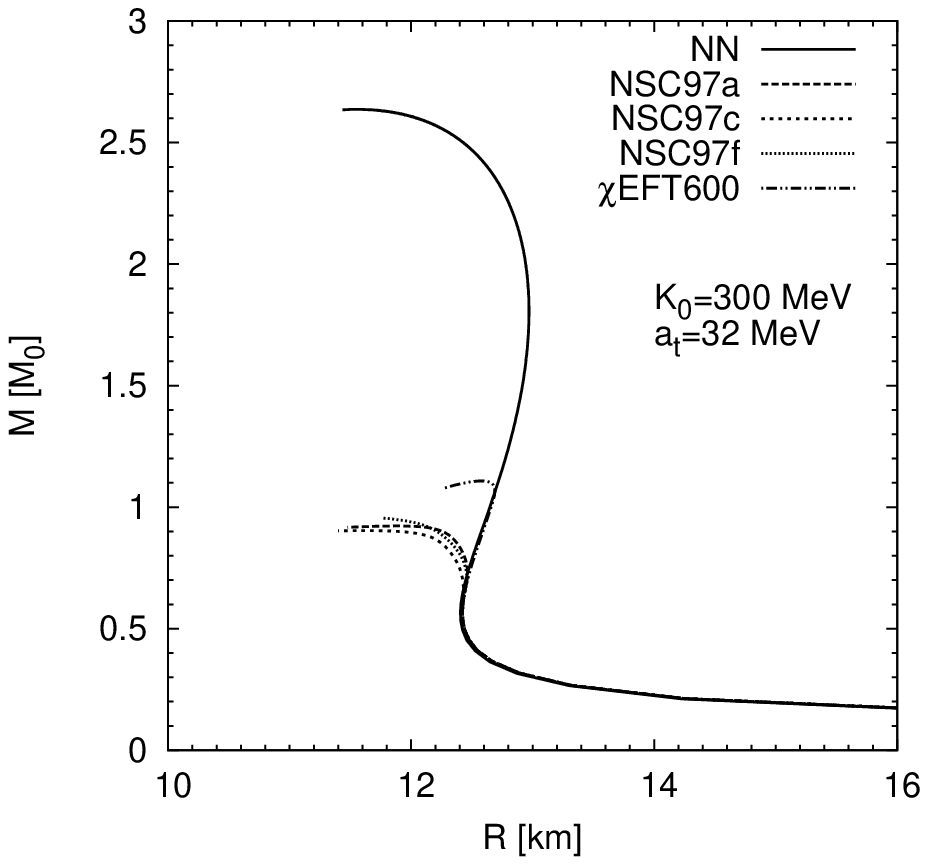}}\hspace{-0.5 cm}
  \caption{Mass-radius relation of a neutron star for symmetry energy
    $a_t=32\;\text{MeV}$ and different $YN$, $YY$ and $\Xi N$
    interactions. For comparison the mass-radius curve obtained for
    the pure $NN$ interaction is also shown. Left panel: soft EoS,
    right panel: stiff EoS. \label{fig:mYYR}}
\end{figure*}

Fig.~\ref{fig:mYYR} shows the mass-radius relation of a NS with
different $YN$, $YY$ and $\Xi N$ interactions is presented. The effect
of the inclusion of the $S=-2$ sector is rather marginal. As is
visible in Fig.~\ref{fig:mYYR} the maximum masses are lower than in
the previous cases which is reasonable since a further degree of
freedom, the $\Xi$ particle, is added. However we note that the
$NSC97$ $YY$ interaction is attractive which is the reason for the
decrease of the allowed maximum NS masses, see
e.g.~\cite{Vidana:2000ew}. A repulsive $YY$ interaction would have an
opposite effect as shown in \cite{Mornas:2004vs}.

It is also interesting to point out that for the $\chi$EFT600
interaction, where the $\Sigma^-$ hyperons appears late, if at all, it
is the $\Xi$ hyperons which takes its pace and influences the maximum
mass of a NS \cite{Glendenning:2000jx, Balberg:1998ug}. For all other
models in which the $\Sigma^-$ appears earlier than the $\Xi$ their
influence on the NS mass is marginal.

\section{Summary and conclusions}
\label{sec:summary}

The main intention of the present work was to study the consequences of available
hyperon-nucleon interactions on the composition of neutrons stars and the 
maximum masses. The analysis was performed in the framework of 
the renormalisation group improved $\ensuremath{V_\text{low k}}$ interaction
deduced from available bare potentials. Since the experimental data base is very limited,
these potentials are not well constrained, in contrast to the nucleon-nucleon case. 

We have determined the threshold densities for the appearence of hyperons in
$\beta$-equilibrated neutron star matter to lowest order in a loop expansion. 
To explore the sensitivity to available $YN$ potentials we have 
constructed single-particle potentials for the $\Lambda$ and
$\Sigma^-$ hyperon in lowest order and deduced the
energy per particle. We have replaced the
pure nucleonic contribution to the energy per particle by an analytic
parameterization. This replacement enables us to vary the
incompressibility $K_0$  and symmetry energy $a_t$ of the purely nucleonic 
EoS and to investigate their influence on the threshold densities of hyperons.
The composition of $\beta$-stable matter has been determined by the
requirement of charge neutrality and $\beta$-equilibrium. The
corresponding threshold densities for various values of $K_0$ and $a_t$ were
evaluated. The most important conclusion is that a more attractive
single-particle potential will decrease the chemical potential and
thus decrease the threshold density of the corresponding hyperon.

We have found that, irrespective of the $YN$ interactions,
incompressibility and symmetry parameter used, hyperons will appear in
dense neutron star matter at densities around $\sim 2\rho_0$. This
inevitably leads to a significant softening of the EoS which in turn results in
smaller maximum masses of a neutron star compared to a purely nucleonic EoS.
Notably, the predicted maximum masses are well below the observed value of
$1.4\;M_{\odot}$, an outcome also known from other works,
e.g.~\cite{Vidana:2000ew, SchaffnerBielich:2002ki,
SchaffnerBielich:2007yq, Mornas:2004vs}. This poses a serious problem.

The softening of the EoS due to hyperons cannot be circumvented by
stiffening the nucleonic EoS, i.e., by increasing $K_0$, since this
will cause hyperons to appear earlier. Changing the high-density
behavior of the symmetry energy dependence or including the $S=-2$
sector does not alter this conclusion either. For more details about
the $S=-2$ sector, in particular $\Xi$ hyperons in dense baryonic
matter see e.g.~\cite{Wang:2005vg, Pal:1999sq, Huber:1997mg,
  Glendenning:1982nc}. This can only mean, that correlations beyond
the one-loop level could be important to stiffen the hyperon
contributions to the EoS. This, however, is not sufficient as
Brueckner-Hartree-Fock calculations indicate \cite{Baldo:1998hd,Baldo:1999rq}. As
has been known for a long time from non-relativistic nuclear many-body
theory, three-body interactions are crucial to yield a stiff nucleonic
EoS. Repulsive three-body forces may also play a role in the hyperon
sector. There is, however, little empirical information available at
present. There is also the possibility of an early onset of
quark-hadron transition to cold quark matter. This might also stiffen
the high-density equation of state.

Clearly, our results pose significant restrictions on any reasonable
equations of state employed in the study of neutron star matter. With the
prediction of a low onset of hyperon appearance it becomes mandatory
to seriously consider strangeness with respect to neutron stars. Eventhough our
predictions for the maximum masses of neutron stars are too low, the
treatment of hyperons in neutron stars is necessary and any approach
to dense matter must address this issue.

\section*{Acknowledgments}
\label{sec:Acknowledgements}

H\DJ\ has been partially supported by the Helmholtz Gemeinschaft under
grant VH-VI-041 and by the Helmholtz Research School for Quark Matter
Studies. BJS acknowledges support by the BMBF under grant 06DA123 and
JW from the Extreme Matter Institute within the Helmholtz Alliance
'Cosmic Matter in the Laboratory'. This work was also supported in
parts by the grant SFB634 of the Deutsche Forschungsgemeinschaft. We
thank R. Alkofer and M. Oertel for a critical reading of the
manuscript.

\end{document}